\documentclass [aps,floats,amsmath,preprint]{revtex4-1}
\usepackage{graphicx}% Include figure files
\usepackage{dcolumn}% Align table columns on decimal point
\usepackage{bm}% bold math
\usepackage{xcolor}
\usepackage{multirow}

\begin{document}

\title{The deformation of a flexible fiber settling in a quiescent viscous fluid}

\author{Benjamin Marchetti$^{1}$, Veronica Raspa$^{2}$, Anke Lindner$^{3}$, Olivia du Roure$^{3}$, Laurence Bergougnoux$^{1}$, \'Elisabeth~Guazzelli$^{1}$, and Camille Duprat$^{2}$}

\affiliation{$^{1}$Aix Marseille Univ, CNRS, IUSTI, Marseille, France\\
$^{2}$LadHyX, Department of Mechanics, CNRS, \'Ecole polytechnique, 91128 Palaiseau, France\\
$^{3}$PMMH, UMR 7636, ESPCI Paris, PSL Research University, Universit\'e Paris Diderot, Universit\'e Pierre et Marie Curie, 10 rue Vauquelin, Paris, France}

\date{\today}

\begin{abstract}
The equilibrium state of a flexible fiber settling in a viscous fluid is examined using a combination of macroscopic experiments, numerical simulations and scaling arguments. We identify three regimes having different signatures on this equilibrium configuration of the elastic filament:  weak and large deformation regimes wherein the drag is proportional to the settling velocity as expected in Stokes flow and an intermediate elastic reconfiguration regime where the filament deforms to adopt a shape with a smaller drag which is no longer linearly proportional to the velocity.
\end{abstract}

\maketitle

\section{\label{sec:intro}Introduction}

The motion of flexible slender bodies in viscous fluids is of fundamental importance in various fields such as biopolymer (e.g. DNA or actin microfilaments) or polymer science \cite{Larson:2005,Graham:2011,Shelley:2016} and pulp and paper or textile engineering \cite{Denn:1980,Lundell:2011}. When these flexible filaments are submitted to a fluid flow or to external forces such as gravity, the interplay between the internal elastic forces of the deformable body and the hydrodynamic forces can lead to complex deformation and motion, which may have strong consequences on their macroscopic transport \cite{lindnershelley2015}. Flow-induced fiber deformation is also a model system to investigate the influence of flexibility on the drag experienced by an object; indeed, the drag is modified since the filament shape becomes a function of its velocity, and can actually be reduced, both at high \cite{alben2002,gosselin2010} and low Reynolds number \cite{CosentinoLagomarsino:2005,alvarado2017}, although the latter regime has received less attention. The present work focusses on one of the simplest flow situation by considering the deformation of a flexible fiber in response to forces which act upon it when settling under gravity in a quiescent viscous fluid.

\begin{figure}
\includegraphics[width=0.55\linewidth]{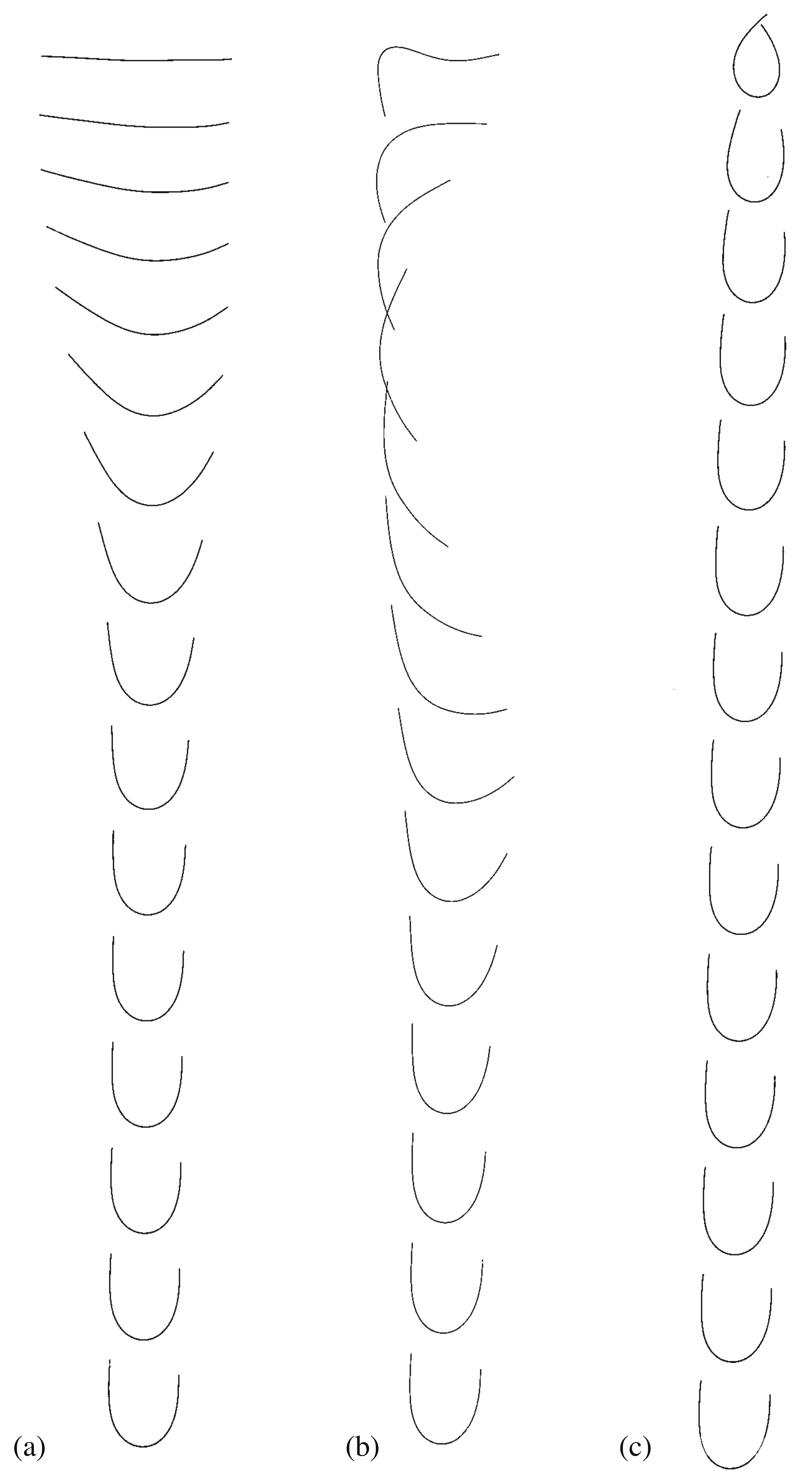}
\caption{Experimental chronophotographies of an elastic filament settling in a viscous fluid for different initial conditions. Note that the fiber is not perfectly homogeneous, resulting in a systematic slight asymmetry of the shape.ADD TIME STEP}
\label{fig:Chrono}
\end{figure}

A long uniform flexible fiber settling in a viscous fluid deforms dynamically in response to the viscous stresses which act upon it. This deformation arises solely because of nonlocal hydrodynamic interactions along the fiber; hydrodynamic interactions with adjacent parts of the fiber are stronger near the middle than near the ends, causing the middle of the fiber to settle faster than its ends.  As a result of this deformation, the flexible fiber experiences a torque which orients it toward a horizontal position, i.e. with its long axis perpendicular to the direction of gravity regardless of its initial configuration, as evidenced in Fig.\,\ref{fig:Chrono}, and eventually adopts a more or less pronounced `U' shape. This has been shown analytically and numerically using slender-body theory \cite{Xu:1994,Li:2013,Shojaei:2015} but also confirmed numerically using discrete modeling of the filament as a string of connected beads interacting by elastic and repulsive forces with different degrees of sophistication \cite{Schlagberger:2005, CosentinoLagomarsino:2005,Llopis:2007,Delmotte:2015}. To the best of our knowledge, no experiments were reported. 

In this paper, we focus on the equilibrium state of a flexible fiber settling in a viscous fluid using a combination of macroscopic experiments, numerical simulations, and scaling arguments. In particular, we identify three different regimes depending on the relative magnitude of gravitational and elastic forces. We explore the signature of these regimes on the shape and velocity of the filament.

\section{\label{sec:Dimensional}Physical mechanisms and scaling arguments}

We consider a fiber, of length $2\ell$ and radius $a$, settling in a quiescent viscous fluid, driven by a gravitational force $F_g$. The fiber experiences a viscous drag such that, at equilibrium, $F_{drag}=F_g$. An elastic fiber deforms in response to the viscous stresses (of magnitude $F_{drag}$, thus at equilibrium $F_g$) and bends along its length in a ``U-shape'', to adopt a typical curvature $1/2\ell$. Balancing the torque applied on the fiber, $F_g \, 2\ell$, and the typical resisting elastic torque, $EI/2\ell$, where $E$ is the Young modulus and $I=\pi a^4/4$ is the second moment of inertia, gives a dimensionless elasto-gravitational number,

\begin{equation}
\mathcal{B}=\frac{F_g (2\ell)^2}{EI}.
\label{eq:B}
\end{equation}

\begin{figure}
\includegraphics[width=0.9\linewidth]{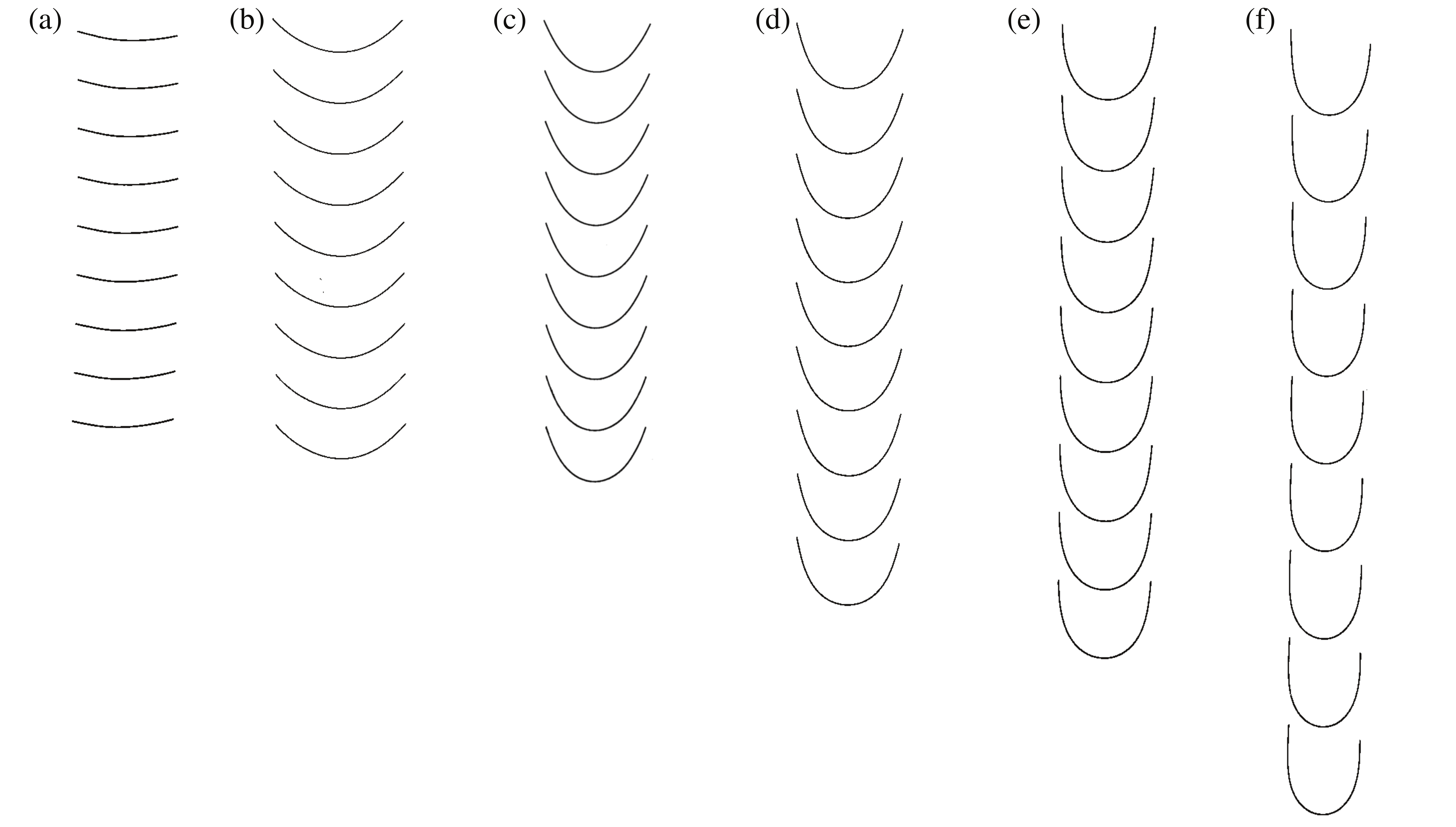}
\caption{Experimental chronophotographies in the stationary state, where the fiber shape and velocity remain constant. All filaments have similar material properties ($a$, $\rho_s$, $EI$) such that the settling velocity of an equivalent rigid filament $U_{\perp}$ is the same. From a) to f), the length of the fiber increases, i.e. $\mathcal{B}$ increases ($\mathcal{B}=57, 111, 207, 222, 329, 439, 549$). The time between successive photos is $10a/U_{\perp}$; a difference in travelled distance thus indicates a difference in velocity.}
\label{fig:ChronoB}
\end{figure}

While the driving force (simply the fiber weight) is known and constant, the expression of the drag force $F_{drag}$ is not known a priori and depends on the shape of the filament, thus on its elastic deformation. The deformation of the fiber, and thus its velocity, are controlled by $\mathcal{B}$ (for a given fiber aspect ratio $\kappa^{-1}=\ell/a$); the relative magnitude of gravitational and elastic forces increases with $\mathcal{B}$ and the filament deformation increases, as exhibited in Fig.\,\ref{fig:ChronoB}.

At small $\mathcal{B}$, the fiber is only weakly deformed, see Fig.\,\ref{fig:ChronoB}\,(a). Its shape remains close to that of a rigid fiber settling perpendicularly to the direction of gravity.  We thus expect the drag to be well approximated by the drag on a rigid filament of same dimension and characteristics. For a rigid filament, the viscous drag is proportional to the filament velocity and length and depends on the filament orientation. A slender fiber, of length $2\ell$ and radius $a$, thus of aspect ratio $\kappa^{-1}=\ell/a$, settling at a velocity $U_{\perp}$ in a fluid of viscosity $\mu$ with its long axis perpendicular to the direction of gravity experiences a drag,
\begin{equation}
\label{eq:coxdrag}
F_{drag\perp}=C_{\perp} \mu U_{\perp} 2\ell,
\end{equation}
with a coefficient $C_{\perp}$ which solely depends on $\kappa^{-1}$ \cite{CoxSlenderBody} and reads
\begin{eqnarray}
C_{\perp} & = & \frac{4\pi}{\ln (4 \kappa^{-1}) -1/2} \quad \mbox{at order} \quad 1/ (\ln \kappa^{-1})^2,\\
 & \simeq & \frac{4\pi}{\ln \kappa^{-1}} \quad \mbox{at leading order} \quad 1/ (\ln \kappa^{-1}).
\end{eqnarray}
Balancing this drag and the gravitational force, $F_g=\Delta \rho g \, \pi a^2 2\ell$ (where $\Delta \rho=\rho_s-\rho_f$ is the density difference between the solid filament and the fluid), yields the settling velocity,
\begin{eqnarray}
U_{\perp}&=&\frac{\Delta\rho g a^2 \left[ \ln (4 \kappa^{-1}) -1/2 \right]}{4\mu} \label{eq:uperp}\\
 & \simeq & \frac{\Delta\rho g a^2 \ln \kappa^{-1}}{4\mu} \quad \mbox{at leading order}.
 \end{eqnarray}
 
%\mathcal{B}$ is our control parameter.
 %that gives the relative magnitude of gravitational and elastic forces
%Deformation of the fiber due to viscous forces, resisted by elasticity, thus given depending on the relative magnitude of gravitational and elastic forces.
%\begin{equation}
%M=EI\frac{\mathrm{d}^2 y}{\mathrm{d}x^2}
%\end{equation}
%where the torque applied on the fiber is $\sim F_gL$ and the moment to bend the fiber over its own length is $EIL/L^2\sim EI/L$. The ratio of these to moments gives the elasto-gravitational number $$\mathcal{B}=\frac{F_gL^2}{EI}$$ that compares the driving force to the elastic resisting torque.

The deformation increases slightly with increasing $\mathcal{B}$, while the velocity remains close to $U_{\perp}$, see Fig.\,\ref{fig:ChronoB}\,(a)-(b). The typical deflection $\delta$ due to viscous forces is given by a balance between the torque applied on the fiber by the viscous drag, $F_{drag}\, 2\ell $, and the bending torque for small deformations $\delta$, $EI\delta/(2\ell)^2$, such that
\begin{equation}
\frac{\delta}{\ell}\propto\mathcal{B}.
\end{equation}
In this regime, the parameter $\mathcal{B}$ can be expressed as 
\begin{equation}
\mathcal{B}=\frac{F_{drag}(2\ell)^2}{EI}=C_{\perp}\frac{\mu U_{\perp} (2\ell)^3}{EI}=C_{\perp}\mathcal{V},
\end{equation}
where we introduce the elasto-viscous number $\mathcal{V}=\frac{\mu U (2\ell)^3}{EI}$ (here with $U=U_{\perp}$).

As $\mathcal{B}$ is further increased, the deflection and the velocity of the fiber increases, see Fig.\,\ref{fig:ChronoB}\,(b)-(e), to reach a large deformation regime where the filament adopts a saturated U shape, i.e. nearly folds onto itself, as shown in Fig.\,\ref{fig:ChronoB}\,(f). In that regime, the deflection is constant,
\begin{equation}
\frac{\delta}{\ell}\simeq 1.
\end{equation}
The two branches of the $U$ are aligned with the flow; we thus expect the drag to be close to that of two rigid fibers of length $\ell$ settling parallel to gravity, i.e. the total drag should approximately be
\begin{equation}
F_{drag}= 2 \, C_{\parallel} \mu U \ell,
\end{equation}
with a coefficient $C_{\parallel}$ which solely depends on $\kappa^{-1}$ as \cite{CoxSlenderBody}
\begin{eqnarray}
C_{\parallel} & = & \frac{2\pi}{\ln(4\kappa^{-1})-3/2} \quad \mbox{at order} \quad 1/ (\ln \kappa^{-1})^2,\\
 & \simeq & 
 \frac{2\pi}{\ln \kappa^{-1}} \quad \mbox{at leading order} \quad 1/ (\ln \kappa^{-1}).
\end{eqnarray}
The drag is smaller than that of a fiber settling perpendicular to gravity, such that the ratio $C_{\perp}/C_{\parallel} \simeq 1.5-1.7$ for $70<\kappa^{-1}<300$ and the settling velocity is higher, since $U_{\parallel}/U_{\perp}= C_{\perp}/C_{\parallel}$.
In this regime, $\mathcal{B}$ can be expressed as
\begin{equation}
\mathcal{B}=C_{\parallel}\mathcal{V}.
\end{equation}

For intermediate values of $\mathcal{B}$, both the deflection and the velocity increase. Indeed, as the filament adopts a more pronounced U-shape, the viscous drag decreases since larger portions of the filament are aligned with the flow. This increase of velocity with flexibility, as the filament deforms to adopt a shape with a smaller drag, is reminiscent of the reconfiguration observed at large Reynolds number \cite{gosselin2010}, but here in a low Reynolds number regime rarely explored. 
In this intermediate reconfiguration regime, the drag force is not known and can not be approximated by either the drag on a perpendicular fiber, or on vertical fibers; it indeed depends on the shape of the fiber, i.e. on an apparent length $\ell_{app}$ that is not the simple length of the fiber as for the cases discussed above but is given by the fiber deformation. This latter deformation is controlled by the typical bending torque, denoted as the stiffness $\mathcal{S}=EI/2\ell$ and the viscous force, i.e. $\ell_{app}=\ell_{app}(\mu U, \mathcal{S})$. Simple dimensional analysis gives a scaling for the apparent length $\ell_{app}\sim(\mathcal{S}/(\mu U))^{1/2}$, and thus implies a new scaling for the drag,
\begin{equation}
F_{drag}\sim \mu U\ell_{app}\sim (\mu U)^{1/2}\mathcal{S}^{1/2}.
\end{equation}
We note that, contrary to the weak or strong deformation regimes where the drag is proportional to $U$, the drag here is proportional to $U^{1/2}$ with a weaker exponent characteristic of a drag reduction regime since the apparent length  depends on $U$. In this regime, the settling velocity, given by $F_{drag}=F_g$, is not a mere constant but varies as 
\begin{eqnarray}
\frac{U}{U_{\perp}}&\sim& \frac{U}{F_{drag}/(\mu2\ell)}\sim\frac{\mu U2\ell}{(\mu U)^{1/2}S^{1/2}},\\ \nonumber
&\sim& \left[\frac{\mu U (2 \ell)^3}{EI}\right]^{1/2}\equiv \mathcal{V}^{1/2}.
\end{eqnarray}
The deflection, given by $F_{drag} 2 \ell \sim EI\delta/(2\ell)^2$, is thus
\begin{equation}
\frac{\delta}{\ell}\sim \mathcal{V}^{1/2}.
\end{equation}
In this regime, we also obtain
\begin{equation}
\mathcal{B} \sim \mathcal{V}^{1/2}.
\end{equation}

We thus have three regimes with different signatures on the equilibrium configuration of the elastic filament. The dimensionless deflection $\delta/\ell$ scales as $\mathcal{V}$ in the weak deformation regime, as $\mathcal{V}^{1/2}$ in the reconfiguration regime, and is constant $\approx 1$ in the large deformation (or saturation) regime. Similarly, the velocity is given by $U/U_{\perp}\simeq 1$ for small deformation,  $U/U_{\perp}\propto\mathcal{V}^{1/2}$ at intermediate deformation, and $U/U_{\perp} \approx 1.6 $ for large deformations. In the following, we present experimental and numerical results to assess these three different scalings.

\section{\label{sec:exp} Experimental techniques}

\begin{figure}
\centering
\includegraphics[width=0.6\linewidth]{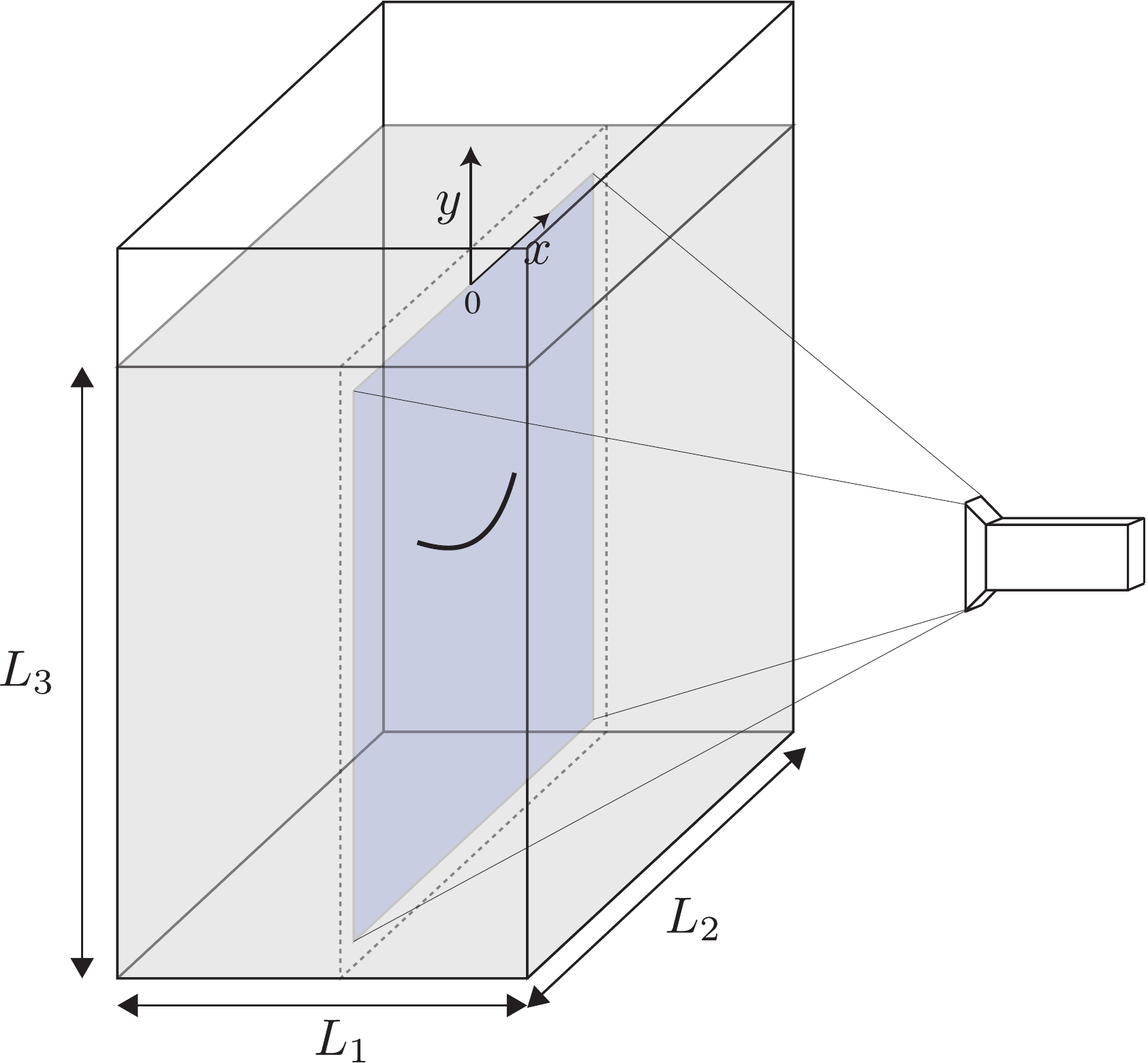}
\caption{Experimental setup. The filament is initially maintained at the center of the tank and aligned with the plane defined by the dotted line. The shaded area corresponds to the plane of view of the camera.}
\label{fig:expsetup}
\end{figure}
Experiments are carried out using the experimental setup depicted in Fig.\,\ref{fig:expsetup}. We use two different containers: a small tank ($L_1=20$ cm, $L_2=20$ cm, $L_3=50$ cm) and a larger tank ($L_1=40$ cm, $L_2=60$ cm, $L_3=80$ cm) to avoid wall effects. The transparent tanks are filled with two different types of liquids (water-based Ucon oil and silicon oil) of various densities and viscosities as indicated in Table \ref{tab:fluid}.

\begin{table}
\begin{ruledtabular}
\begin{tabular}{cccc}
 & Fluid 1 & Fluid 2 & Fluid 3\\
\hline
Mixture & Silicon Oil & $50\%$ water + 50$\%$ Ucon Oil$^{\mbox{\scriptsize\textregistered}}$& $60\%$ water + 40$\%$ Ucon Oil$^{\mbox{\scriptsize\textregistered}}$ \\
\hline 
$\rho_f$(kg.m$^{-3})$& 970  & 1074 & 1061  \\
\hline
$\mu_f$(Pa.s)& 0.97 & 0.96 & 0.30  \\

\end{tabular}
\caption{\label{tab:fluid}Main characteristics of the fluids used in the experiments.}
\end{ruledtabular}
\end{table}

Fibers are fabricated from a silicon-based elastomer (Zermak Elite double 8) molded in capillary tubes. The  density is tuned by adding iron powder in different proportions, which also slightly modifies the Young modulus. The Young modulus is determined for each solution by standard traction measurements. The properties of the elastomeric filaments are highly sensitive to storage conditions, and may vary depending on the fluid they are stored in. In particular, the elastomer swells in silicon oil; new filaments are casted every day to ensure constant properties (we have verified that, when immersed in silicon oil, the filaments properties remained unchanged for several days, but we imposed a shorter time of use, typically 24 hours, to ensure reproducibility). In water-based fluids, no swelling is observed and filaments can be extracted and kept in air. Under these conditions, aging and solvent evaporation may cause changes in properties from fabrication (in particular, hardening can be observed). The properties of the filaments are then measured at their time of use in the experiments. The properties of the filaments are presented in Table \ref{tab:fibers}.

\begin{table}
\begin{ruledtabular}
\begin{tabular}{cccccccccc}
 \multicolumn{5}{c}{Batch P} &  \multicolumn{5}{c}{Batch M}\\
Fe $\% w/W$ & $\rho_s$(kg.m$^{-3}$) & $E$ (kPa) & $a$ ($\mu$m) & $\mathcal{B}$(-) & Fe $\% w/W$ & $\rho_s$ (kg.m$^{-3}$) & $E$ (kPa) & $a$ ($\mu$m) & $\mathcal{B}$(-)\\
\hline 
\multirow{2}{*}{10} &  \multirow{2}{*}{1166} & \multirow{2}{*}{218.3} & 128.9 & 16-1090 & \multirow{2}{*}{10} &  \multirow{2}{*}{1166} & \multirow{2}{*}{198} & 138 & 373-430\\
 \cline{4-5} \cline{9-10}
 & & & 229.6 & 60-1030& & & & 232 & 65\\

 \hline
 \multirow{2}{*}{18} &  \multirow{2}{*}{1254} & \multirow{2}{*}{243.0} & 128.9 & 22-1410 &\multirow{2}{*}{20} &  \multirow{2}{*}{1295} & \multirow{2}{*}{220} & 140 & 221-555\\
 \cline{4-5} \cline{9-10}
 & & & 229.6 & 33-445& & & & 232 & 187-203\\
  
 \hline
  \multirow{2}{*}{20} &  \multirow{2}{*}{1295} & \multirow{2}{*}{251.5} & 128.9 & 24-1560&\multirow{2}{*}{30} &  \multirow{2}{*}{1450} & \multirow{2}{*}{1100} & 500 & 40\\
 \cline{4-5} \cline{9-10}
 & & & 229.6 & 57-550& & & & 285 & 88\\

 \hline
 &   & &  & &\multirow{2}{*}{40} &  \multirow{2}{*}{1617} & \multirow{2}{*}{226} & 137 & 120-286\\
 \cline{9-10}
 & & & & & & & & 232 & 56-110\\
 
\end{tabular}
\caption{\label{tab:fibers}Main characteristics of the filaments used in the experiments. The length of the filament varies between $2\,\mathrm{cm} <2\ell< 8\,\mathrm{cm}$.}
\end{ruledtabular}
\end{table}

Two independent sets of data have been collected: filaments of batch~P settling in fluid 1 inside the large tank, and filaments of batch~M settling in fluids 2 and 3 inside the smaller tank. The elasto-gravitationnal number, $\mathcal{B}$, spans a large range to cover all regimes, $60\lesssim\mathcal{B}\lesssim1200$. The Reynolds number, $Re= U \ell \rho_f/\mu$, is always smaller than $0.01$ for experiments done in fluids 1 and 2 but can reach $0.2$ for the less viscous fluid 3. 

\begin{figure}
\centering
\includegraphics[width=0.98\linewidth]{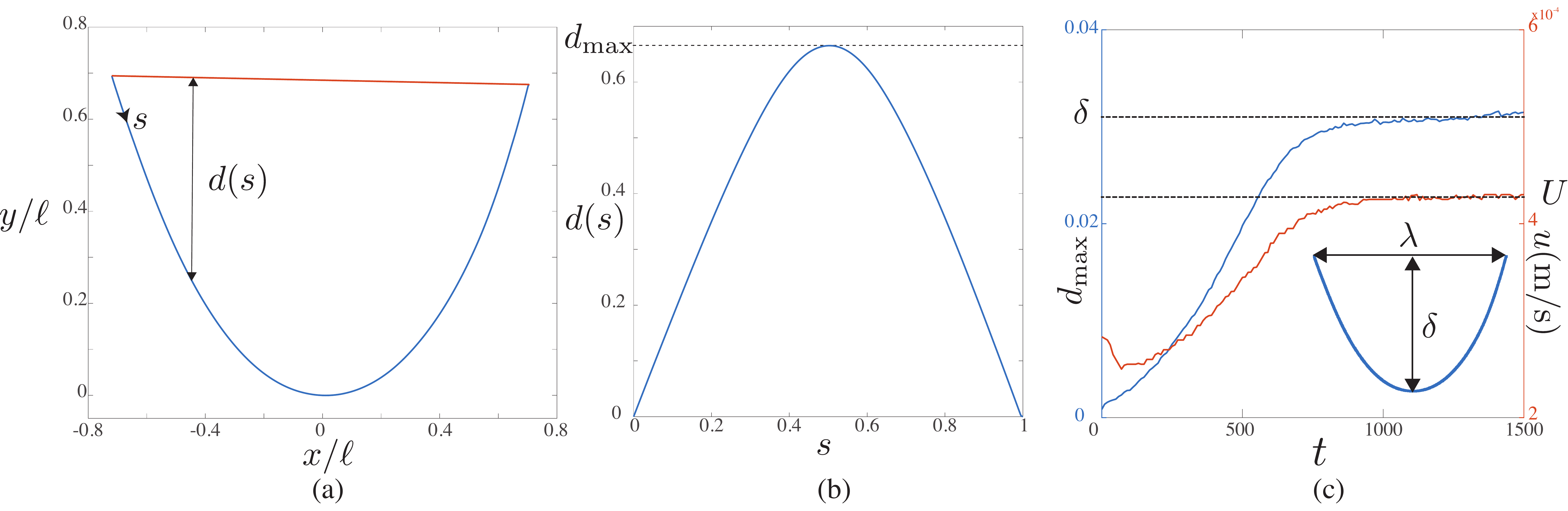}
\caption{Image analysis (typical profiles obtained experimentally): the instantaneous shape $y(x)$ of the filament is extracted from the images (a), the deformation $d(s)$ is measured along the arc length (b) and the maximum deflection $d_{max}$, as well as the velocity $u$ are followed with time (c). The final stationary shape (inset in (c)) is symmetric, and has a maximum deflection $\delta$ and an end-to-end distance $\lambda$.}
\label{fig:expana}
\end{figure}

A pair of reversed action mechanical tweezers placed at the center of the container enables to hold and to release the fiber while minimizing the surface of contact with the filament and any undesired torsion or tension. A specially designed pool having a rectangular opening at its bottom has been alternatively used in the smaller tank experiments. The filaments are thus released in a plane perpendicular to the wall of the container (indicated by the dotted area in Fig.\,\ref{fig:expsetup}). The shape and position of the filament are recorded (typically at 0.1-0.5 fps) with a high-resolution digital camera having a wide-angle lens in order to image the entire settling dynamics (see the shaded area in Fig.\,\ref{fig:expsetup}). We verify that the settling is planar with a second camera placed perpendicularly to the observation plane.

The shape $y(x)$ of the fiber is extracted thanks to standard image processing functions, see a typical example in Fig.\,\ref{fig:expana}\,(a). On this shape, we measure the deflection $d(s)$ along the length of the fiber ($s$ denotes the arc length) and we extract the maximum deflection $d_{max}$ on each shape, see Fig.\,\ref{fig:expana}\,(b). In addition, we record the position of the center of mass of the filament at each time step in order to evaluate the instantaneous vertical velocity $u(t)$. We then follow the evolution of $d_{max}$ and $u$ as the filaments settles, see Fig.\,\ref{fig:expana}\,(c). The deflection increases to reach a constant value $\delta$ while simultaneously the settling velocity saturates at a constant value $U$, indicating that the filament has reached its equilibrium configuration. This procedure is automatized using an in-house custom code, and for each experiment we record the stationary values of the velocity, $U$, the maximum deflection, $\delta$, and the end-to-end distance, $\lambda$. In the following, we report averaged measurements of these quantities over several runs (typically 4-5 for batch M and 1-2 for batch P) and the uncertainties on these measurements are taken as the standard deviations over these runs. 

It is important to stress that there are some unavoidable difficulties in performing these experiments at low Reynolds numbers that result in scatter in the data. First, a small convection current due to a weak thermal gradient is always present across the tanks. This convection current which is typically of the order of $1 \mu$m/s affects the trajectory of the flexible filament and its velocity, in particular in water-based fluids. Second, the bottom and side walls of the tanks may also influence the dynamics of the filament by slowing them down (in particular in the smaller tank, i.e. for experiment with batch M). These effects result in an increased uncertainty on the settling velocity, while not affecting the shape of the filament, as the latter is determined by the velocity difference between the settling filament and the fluid.

The two experimental set-ups are thus complementary. The use of water-based fluids is more amenable for a larger number of experiments to be performed as individual filaments can be used repeatedly, whereas the use of silicon oil in a large tank results in a better resolution on the settling velocity. We will discuss all three sets of data obtained with these two set-ups throughout the paper.

\section{\label{sec:modeling} Theoretical modeling}

\subsection{\label{sec:slender} Slender-body model}

This model, first proposed by Xu and Nadim \cite{Xu:1994} and later revisited in \cite{Li:2013,stoneduprat2015}, is based on slender-body theory \cite{CoxSlenderBody} and considers the weak deflection of a long filament. It is therefore limited to the regime of small values of $\mathcal{B}$. 

The hydrodynamic force acting on a long rigid filament settling in a viscous fluid per unit length of the centerline can be expressed as, 
\begin{equation}
\label{eq:ForcePerLengthCylinder1}
f^H(x) =  2\pi \mu U_{\perp} \left \{-\frac{2}{\ln \kappa} -\frac{1}{\left (\ln \kappa\right )^2}\left [1+2\ln 2+\ln \left (1-\left (\frac{x}{\ell}\right )^2\right )\right ]
+
O\left (\frac{1}{\left (\ln \kappa \right )^3} \right)\right \}.
\end{equation}
While this viscous force is constant along the length at first order, at second order the force increases near the ends of the filament due to non-local effects, leading to the U-shape observed in the experiments. 

The velocity is equal to $U_{\perp}$ and is given by the balance of the viscous and gravitational forces,
\begin{eqnarray}
\Delta \rho g \, \pi a^2 2\ell= \int_{-\ell}^\ell f^H(x)~{\rm d}x &= &\frac{8\pi \mu U_{\perp}\ell}{\ln 4\kappa^{-1}-1/2},\\
&\simeq & \frac{8\pi \mu U_{\perp}\ell}{\ln \kappa^{-1}} \quad \mbox{at leading order},
 \label{eq:velocity}
\end{eqnarray}
as defined earlier in Sec.\,\ref{sec:Dimensional}. 

For small deformation, the force applied on the filament remains equal to Eq.\,(\ref{eq:ForcePerLengthCylinder1}) and the deformation is given by Euler-Bernoulli beam theory \cite{Xu:1994}. The stationary shape of the filament $y(x)$ is then given by 
\begin{equation}\label{eq:beam}
EI\frac{\mathrm{d}^4 y}{\mathrm{d}x^4}=f(x),
\end{equation}
where $f$ is the net force per unit length acting on the fiber, i.e.
\begin{equation}
f(x)=f^{H}(x) - \frac{1}{2\ell}\int_{-\ell}^\ell f^H(x)~{\rm d}x.
\end{equation}
The deformation is given by the second order $1/(\ln \kappa^{-1})^2$ terms. Replacing $f(x)$ in Eq.\,(\ref{eq:beam}), and rescaling with $X=x/\ell$ and $Y=y/y_0$, leads to
\begin{equation}\label{eq:shape_adim}
 \frac{\mathrm{d}^4Y}{\mathrm{d}X^4}=2\ln 2 -2 - \ln \left(1-X^2\right),\\ 
\end{equation}
with a typical deflection,
\begin{eqnarray}
y_0 & = & \frac{2\pi\mu U_{\perp}\ell^4}{EI(\ln \kappa)^2},\\
&\simeq&\frac{\mathcal{B}\ell}{16 \ln\kappa^{-1}} \quad \mbox{at leading order}.
\end{eqnarray}
We can solve (\ref{eq:shape_adim}) with boundary conditions $Y(0)=0, \, Y'(0)=0$ at the center of the beam and $Y''(1)=Y''(-1)=0, \, Y'''(1)=Y'''(-1)=0$ at the free ends,  to obtain the profile 
\begin{equation}\label{eq:profcyl_adim}
Y(X)=\frac{1}{24}\left[X^2+\frac{13}{6}X^4+2\ln 2(6X^2+X^4)- (X-1)^4\ln(1-X)-(X+1)^4\ln(1+X)\right].
\end{equation}
Note that this solution is given in \cite{Xu:1994} with a typo (3/16 instead of 13/6) and in \cite{Li:2013} with a different typo, while a correct form is given in \cite{stoneduprat2015}.
The maximum deflection $\delta$ is given by $Y(\lvert X \lvert = 1)$. For the stationary shape given by Eq.\,(\ref{eq:profcyl_adim}), we find $Y(\lvert X \lvert = 1)=0.0074$ which gives for the maximal amplitude of deformation
\begin{equation}
\label{deltacyl}
\frac{\delta}{\ell}=0.0074~ \frac{y_0}{\ell} \approx \frac{0.0046}{\ln\kappa^{-1}}\mathcal{B}.
\end{equation}
In this small deformation regime, we recover the linear scaling $\delta/\ell \propto \mathcal{B}$ with $U=U_{\perp}$.
Note that there is also a dependence on aspect ratio, $\kappa^{-1}$.

\subsection{\label{sec:bead-spring} Bead-spring model}

\begin{figure}
\centering
\includegraphics[width=0.5\linewidth]{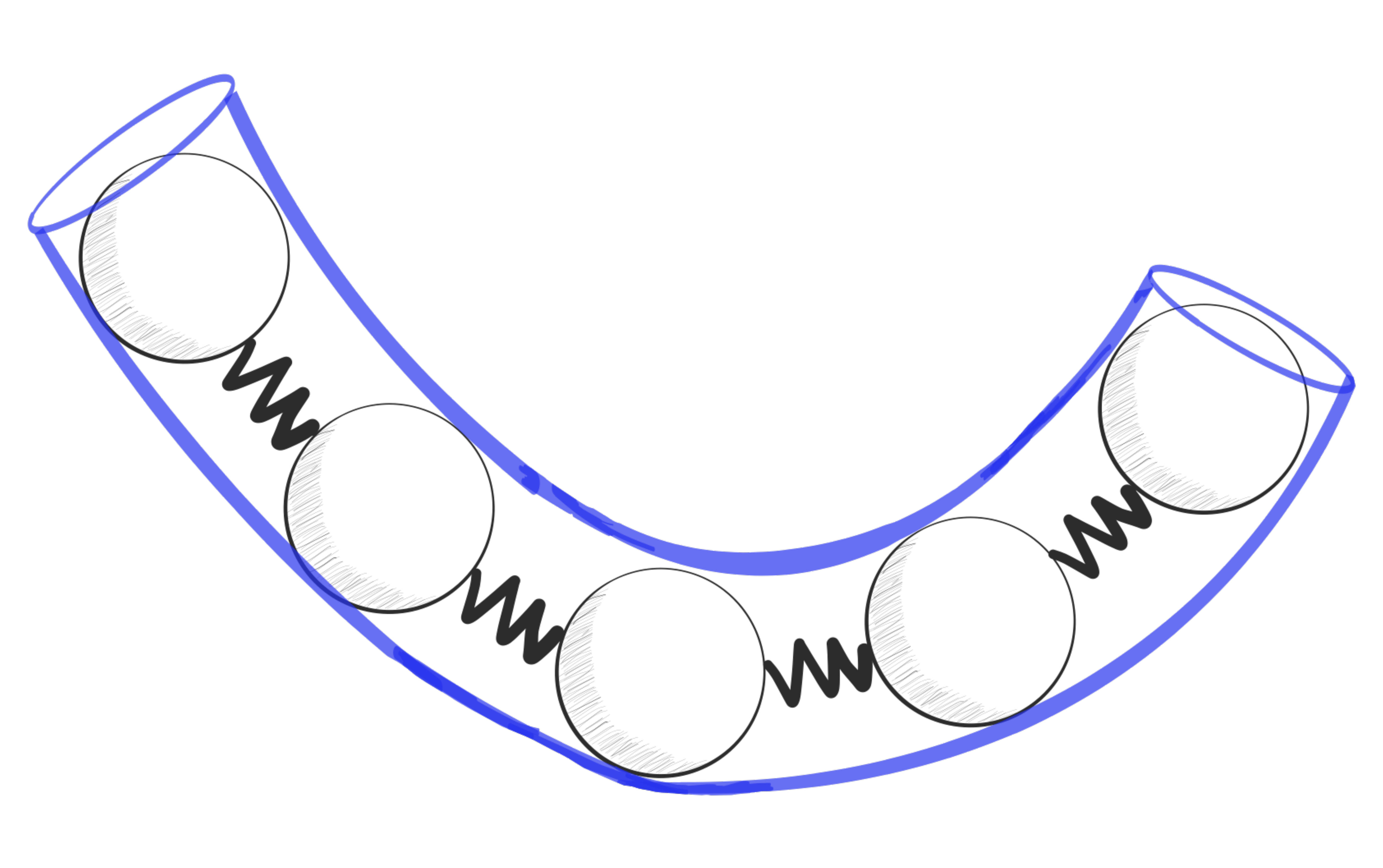}
\caption{Sketch of the bead-spring model wherein the filament is modeled as a chain of spherical beads connected by spring.}
\label{fig:bsm_sketch}
\end{figure}

To tackle all regimes of $\mathcal{B}$ (i.e. in particular to encompass the large-$\mathcal{B}$ range which is not described by the slender-body model presented in the previous Sec.\,\ref{sec:slender}), we also choose to use a discrete modeling wherein the filament is treated as a chain comprising $N=\kappa^{-1}(=\ell/a)$ spherical beads of radius $a$ connected by springs, as illustrated in Fig.\,\ref{fig:bsm_sketch}. This model has been used extensively in the literature with different degrees of approximation and refinement \cite{Schlagberger:2005, CosentinoLagomarsino:2005,Llopis:2007,Delmotte:2015}.

The beads interact through multi-body hydrodynamic interactions and elastic forces. The force balance on each sphere (neglecting particle inertia) can be framed into a mobility problem. The velocity $\dot{\mathbf{r}}^{\alpha}$ of the sphere $\alpha$ located at position $\mathbf{r}^{\alpha}$ and interacting with other spheres $\beta$ located at position $\mathbf{r}^{\beta}$ is given by
\begin{equation}
\dot{r}_{i}^{\alpha} = \sum_{\beta} \mathcal{M}_{ij}^{\alpha\beta}  \left( F_{j}^{\beta} - \frac{\partial \mathcal{U}}{\partial r_{j}^{\beta}} \right),
\label{eq:bsc_model}
\end{equation}
where  $\mathcal{U}$ is the elastic potential and  $\mathbf{F}^{\beta}$ is the external force due to gravity on each particle which is precisely balanced by the Stokes drag and is thus  $=6 \pi \mu a \mathbf{U_S}$ where $\mu$ is the fluid viscosity and $\mathbf{U_S}$ the Stokes settling velocity. Following the approach developed in \cite{Schlagberger:2005}, the elastic potential $\mathcal{U}$ stems from a discrete version of the worm-like chain model and is written as
\begin{equation}
\mathcal{U} = \sum_{\gamma} \left[ a \, S \left( \frac{\mathbf{r}^{\gamma,\gamma+1}}{2a} -1 \right)^2 +  \frac{B}{2a}  \left( 1 - \cos \theta^{\gamma,\gamma+1} \right)  \right],
\label{eq:bsc_U}
\end{equation}
where, for an isotropic elastic cylinder, the stretching and bending moduli are $S=E \pi a^2$ and $B=E \pi a^4/4$, respectively, with $E$ the Young modulus and where $\mathbf{r}^{\gamma,\gamma+1}=\mathbf{r}^{\gamma+1}-\mathbf{r}^{\gamma}$ is the distance between neighboring spheres $\gamma$ and $\gamma+1$ and $\theta^{\gamma,\gamma+1}$ the angle between neighboring bonds $\mathbf{r}^{\gamma,\gamma-1}$ and $\mathbf{r}^{\gamma,\gamma+1}$.
The  mobility tensor $\mathcal{M}_{ij}^{\alpha\beta}$ is the distance-dependent tensor which accounts for hydrodynamic interactions between spheres. We choose to use the Rotne-Prager-Yamakawa tensor,
\begin{equation}
\mathcal{M}_{ij}^{\alpha\beta}  = \frac{1}{6 \pi \mu a}  \left\{  \frac{3}{4} \left[\frac{\delta_{ij}}{\frac{r^{\alpha,\beta}}{a}} +  \frac{\frac{r^{\alpha,\beta}_i r^{\alpha,\beta}_j}{a^2}}{{(\frac{r^{\alpha,\beta}}{a})}^3} \right]  + \frac{3}{2} \left[\frac{\delta_{ij}}{3\,{(\frac{r^{\alpha,\beta}}{a})}^3} - \frac{\frac{r^{\alpha,\beta}_i r^{\alpha,\beta}_j}{a^2}}{{(\frac{r^{\alpha,\beta}}{a})}^5} \right]     \right\},
\label{eq:RPY}
\end{equation}
which takes into account the hydrodynamic interaction between particles up to order $O(\frac{a}{r^{\alpha,\beta}})^{3}$ where $\mathbf{r}^{\alpha,\beta}=\mathbf{r}^{\beta}-\mathbf{r}^{\alpha}$ is the distance between the spheres $\alpha$ and $\beta$ with $r^{\alpha,\beta}= |\mathbf{r}^{\alpha,\beta}|$.
The self mobility is chosen as the Stokes mobility,
\begin{equation}
\mathcal{M}_{ij}^{\alpha\alpha}  = \frac{\delta_{ij}}{6 \pi \mu a}.  
\label{eq:RPYself}
\end{equation}

Other approximations for the mobility tensor could be considered. The leading order Stokeslet approximation used in particular in \cite{CosentinoLagomarsino:2005} corresponds to keeping only the first term (inside the square brackets) on the right-hand side of equation\,(\ref{eq:RPY}). A fuller Rotne-Prager-Yamakawa tensor which provides a regularization for $r^{\alpha,\beta}<2a$ and is positive definite for all the particle configurations can also be used  \cite[see e.g.][]{Wajnryb:2013}. No significant differences are seen between this later fuller tensor and the Rotne-Prager-Yamakawa tensor described by Eqs.\,(\ref{eq:RPY}) and\,(\ref{eq:RPYself}). Discrepancies arise with the Stokeslet approximation for high values of $\mathcal{B}$.

Eq.\,(\ref{eq:bsc_model}) governing the time evolution of the sphere positions can be made dimensionless by using $a$ as the length-scale and $6 \pi \mu a U_S$ as the force-scale, which reads
\begin{equation}
\hat{\dot{r}}_{i}^{\alpha} = \sum_{\beta} \hat{\mathcal{M}}_{ij}^{\alpha\beta}  \left( \hat{F}_{j}^{\beta} - \mathcal{E} \frac{\partial \hat{\mathcal{U}}}{\partial \hat{r}_{j}^{\beta}} \right).
\label{eq:ND_bsc_model}
\end{equation}
Eq.\,(\ref{eq:ND_bsc_model}) exhibits the dimensionless parameter $\mathcal{E} = \frac{E \pi a ^2}{6 \pi \mu a U_S}$. Care should be taken when relating this dimensionless number $\mathcal{E}$ to the elasto-gravitational number $\mathcal{B}$, defined by  Eq.\,(\ref{eq:B}), of a real filament. An important point, which may have been left unnoticed in previous numerical work using the bead-spring model since no comparison with experiments was intended, is that the volume of the modeled object should be that of a filament, as illustrated by the (blue) sheath around the chain of beads in Fig.\,\ref{fig:bsm_sketch}. This means that $\mathcal{B}=32 \Delta \rho g \ell^3/ E a^2 \equiv 24 \kappa^{-3}/\mathcal{E}$. 

Integration of the positions of each sphere is performed using an explicit Runge-Kutta method of order (4)5 (the `dopri5' integrator of the `ode' solver in Python). The Python code is given as supplementary material. Similar methods as those used in the experiments (described at the end of Sec.\,\ref{sec:exp}) are applied to determine the stationary values of the filament velocity, $U$, the maximum deflection, $\delta$, and the end-to-end distance, $\lambda$.

\section{\label{sec:results} Results and comparisons}

\subsection{\label{sec:shapes} Final shape}

\begin{figure}
\centering
\includegraphics[scale= 0.16]{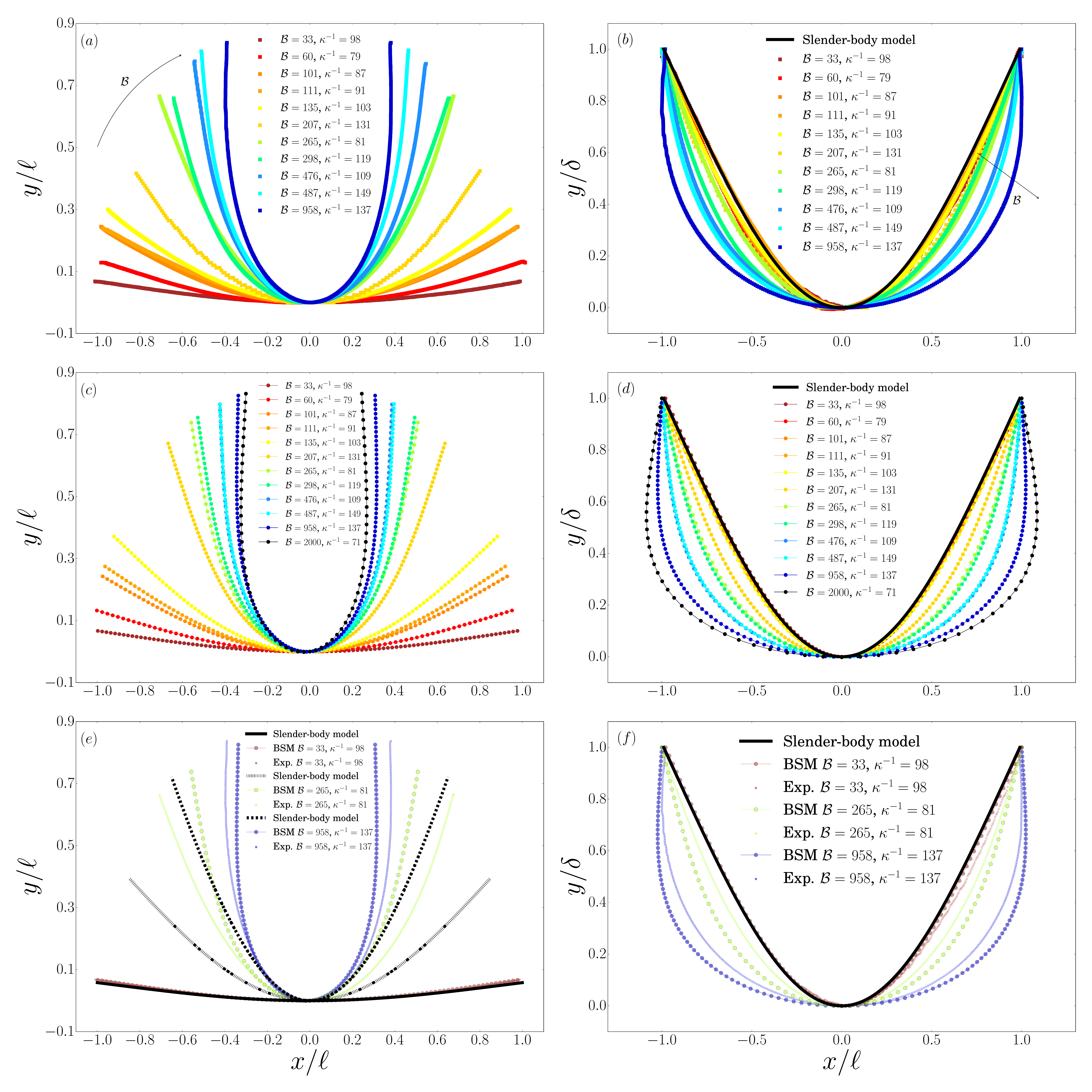}

\caption{Final shapes of the filaments, $\frac{y(x/\ell)}{\ell}$, (left column) and same but rescaled by the maximum amplitude, $\frac{y(x/\ell)}{\delta}$, (right column) for experiments (top), graphs (a) and (b), bead-spring modeling noted BSM (middle), graphs (c) and (d), and selected comparison between experiments and numerical predictions (bottom), graphs (e) and (f).}
\label{fig:shapes}
\end{figure}

Experiments are performed for various $\mathcal{B}$ ($30\lesssim\mathcal{B}\lesssim1000$) and $\kappa^{-1}$ ($70\lesssim\kappa^{-1}\lesssim300$). A selection of the obtained final shapes is presented in Fig.\,\ref{fig:shapes}\,(a). We rescale these profiles with the maximum amplitude $\delta$ in Fig.\,\ref{fig:shapes}\,(b). As $\mathcal{B}$ increases, the amplitude of the deformation increases. Whereas the filament adopts a ``V'' shape at weak deformation, it achieves a ``U'' shape at stronger deformation as its ends become aligned vertically. All the profiles for $\mathcal{B} \leq 200$ collapse onto a single curve in agreement with the prediction (the black line) given by Eq.\,(\ref{eq:profcyl_adim}), i.e. the profile derived by the slender-body model \cite{Xu:1994} presented in Sec.\,\ref{sec:slender}. For larger values, the shape significantly deviates from this profile. A similar trend is recovered in the numerical bead-spring modeling, see Fig.\,\ref{fig:shapes}\,(c)-(d). These numerical simulations enable the exploration of a larger range of $\mathcal{B}$, and in particular to extend the analysis to the high-$\mathcal{B}$ regime (up to $\mathcal{B} =10^4$) where experiments are hardly amenable at a macroscopic scale. As $\mathcal{B}$ is increased, the predicted shape evolves from a ``V'' to a ``U'' shape as seen in the experimental observations. For the higher $\mathcal{B}$ explored, the filament can even reach a highly deformed ``horseshoe'' shape. For intermediate values of $\mathcal{B}$, the shape evolves between the limit (the black line) given by the slender body theory, i.e. the ``V'' shape profile given by Eq.\,(\ref{eq:profcyl_adim}), and the highly deformed ``horseshoe'' shape. Experiments and simulations compare favorably as depicted in Fig.\,\ref{fig:shapes}\,(e)-(f). The largest discrepancy is observed in the intermediate regime, e.g. for $\mathcal{B}=265$, as will be discussed in the following paragraphs.

Some understanding of the obtained shapes can be inferred from examining the combined effect of hydrodynamic and elastic forces on the chain of spheres. As pointed in Sec.\,\ref{sec:intro}, the central spheres settle faster than the end spheres since they experience stronger hydrodynamics disturbances due to the motion of the other spheres. The bending of the filament (of uniform thickness) is thus caused by non-local hydrodynamic interactions. The equilibrium shape is obtained by the balance of the hydrodynamic and elastic forces on the chain. Strong elastic forces, i.e. small $\mathcal{B}$, maintain a weakly curved shape, while smaller elastic forces, i.e. larger $\mathcal{B}$, are less likely to resist the hydrodynamic forces leading to an increased deformation. The highly deformed ``horseshoe'' shape is reminiscent of the subsequent evolution of the chain of spheres without elastic restoring forces: the central spheres settle faster leaving behind the end spheres which then come closer and start to settle faster and to catch up the central spheres, leading finally to a toroidal circulation of the cluster of particles, see e.g. chapter 6 in \cite{GuazzelliMorris2012}. 

\begin{figure}
\includegraphics[width=0.6\linewidth]{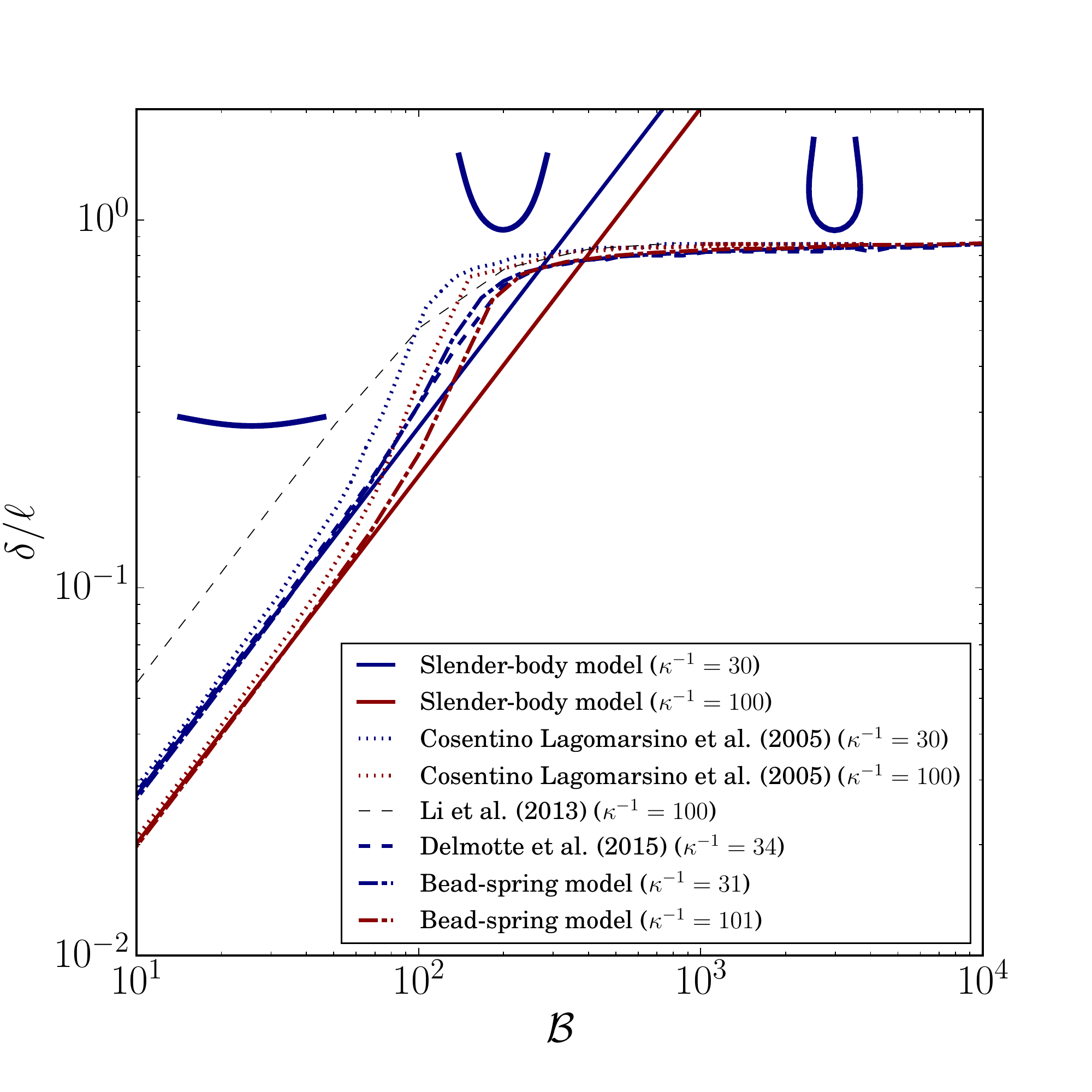}
\caption{Scaled maximum amplitude, $\delta/\ell$, versus $\mathcal{B}$. Comparison of the slender-body and bead-spring models presented in Sec.\,\ref{sec:modeling} with numerical results from previous models \cite{Li:2013,CosentinoLagomarsino:2005,Delmotte:2015} at similar $\kappa^{-1}=30$ and $100$.}
\label{fig:amplitude_biblio}
\end{figure}

The shape of the stationary filament is characterized by its maximum amplitude $\delta$ as well as its end-to-end distance $\lambda$. Before embarking in a detailed comparison between the experiments and the predictions of the models presented in Sec.\,\ref{sec:modeling}, these models are compared to previous numerical results collected in the literature for the maximum deflection $\delta$ in Fig.\,\ref{fig:amplitude_biblio}, for two values of $\kappa^{-1}(=30$ and $100$). In particular, we focus the comparison on two previous bead-spring models, that of \cite{CosentinoLagomarsino:2005} using a Stokeslet approximation and that of \cite{Delmotte:2015} using a Gear model based on a no-slip condition between the beads and ensuring a non extensibility condition for the filament. An important point to mention is that, in calculating $\mathcal{B}$ in these bead-spring models, one must use the volume of the object as a filament and not as a chain of beads (there is a factor 3/2 difference) as explained at the end of Sec.\,\ref{sec:bead-spring}.
At small $\mathcal{B}$, the predictions coming from all the models recover the linear evolution given by the slender-body theory given by Eq.\,(\ref{deltacyl}) \cite{Xu:1994}. The numerical models also recover the same saturation at large $\mathcal{B}$, which, as expected, cannot be captured by the slender-body theory valid only for small deflections. Note that this is not exactly a saturation, as the shape of the filament continues to evolve slightly and the deflection slowly increases while remaining close to $\delta \simeq 0.85 \,\ell$. Note also that a metastable ``W'' shape can be reached for very large values of $\mathcal{B}$. This ``W'' configuration observed also in \cite{CosentinoLagomarsino:2005,Delmotte:2015} is unstable and, after some transient time, the filament rotates and eventually adopts a final ``horseshoe'' shape, see the movie given as supplementary material (for 51 beads and $\mathcal{E}=120$). This transient shape is only observed for large values of $\mathcal{B}>5000$ which can not be attained experimentally; indeed, the ``W'' configuration is never observed in the experiments . The main difference observed between the models is visible in the intermediate (reconfiguration) regime, where the bead-spring models deviate from the linear variation predicted by the slender body model before the saturation regime. The present bead-spring model (introduced in Sec.\,\ref{sec:bead-spring}) agrees well with the Gear model of \cite{Delmotte:2015}, even though they consider different extensibility conditions. The bead-spring  model of \cite{CosentinoLagomarsino:2005} using a Stokeslet approximation shows larger deviations to the linear variation than these two later models which consider a higher degree of approximation for the hydrodynamic interactions. 
To be comprehensive, we have also reported on the graph of Fig.\,\ref{fig:amplitude_biblio} the results of \cite{Li:2013} for a filament of non-uniform thickness (i.e. having an ellipsoid form).  A linear variation with $\mathcal{B}$ is observed at small $\mathcal{B}$ but with a larger relative amplitude than for the uniform filaments considered in the bead-spring models. Interestingly, saturation is observed at the same $\delta \simeq 0.85\,\ell$. Note that this model does not present the deviation from the linear variation observed in the intermediate regime with the bead-spring models.
To conclude, these comparisons appraise the validity of the models used in the present work (described in Sec.\,\ref{sec:modeling}). The experimental results are presented against these two models in the following. 

\begin{figure}
\centering
\includegraphics[width=1\linewidth]{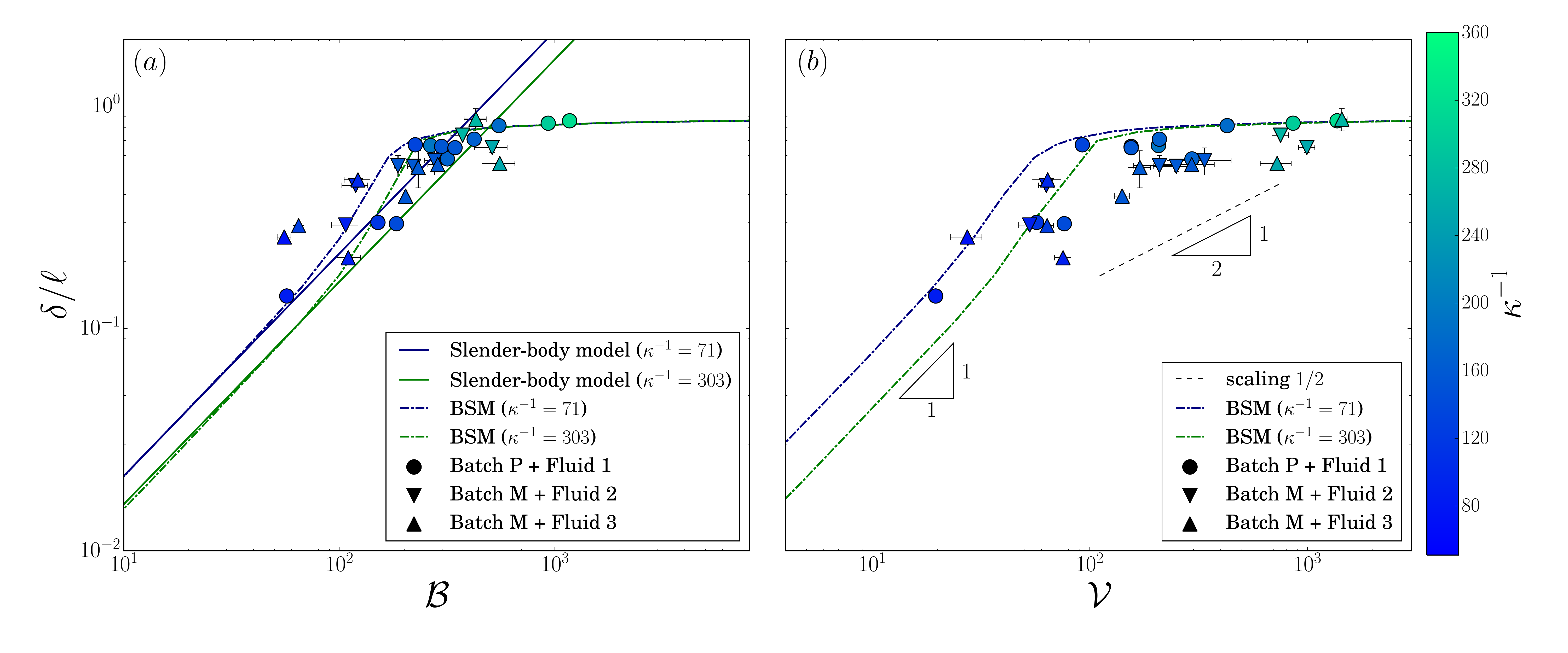}\\
\includegraphics[width=1\linewidth]{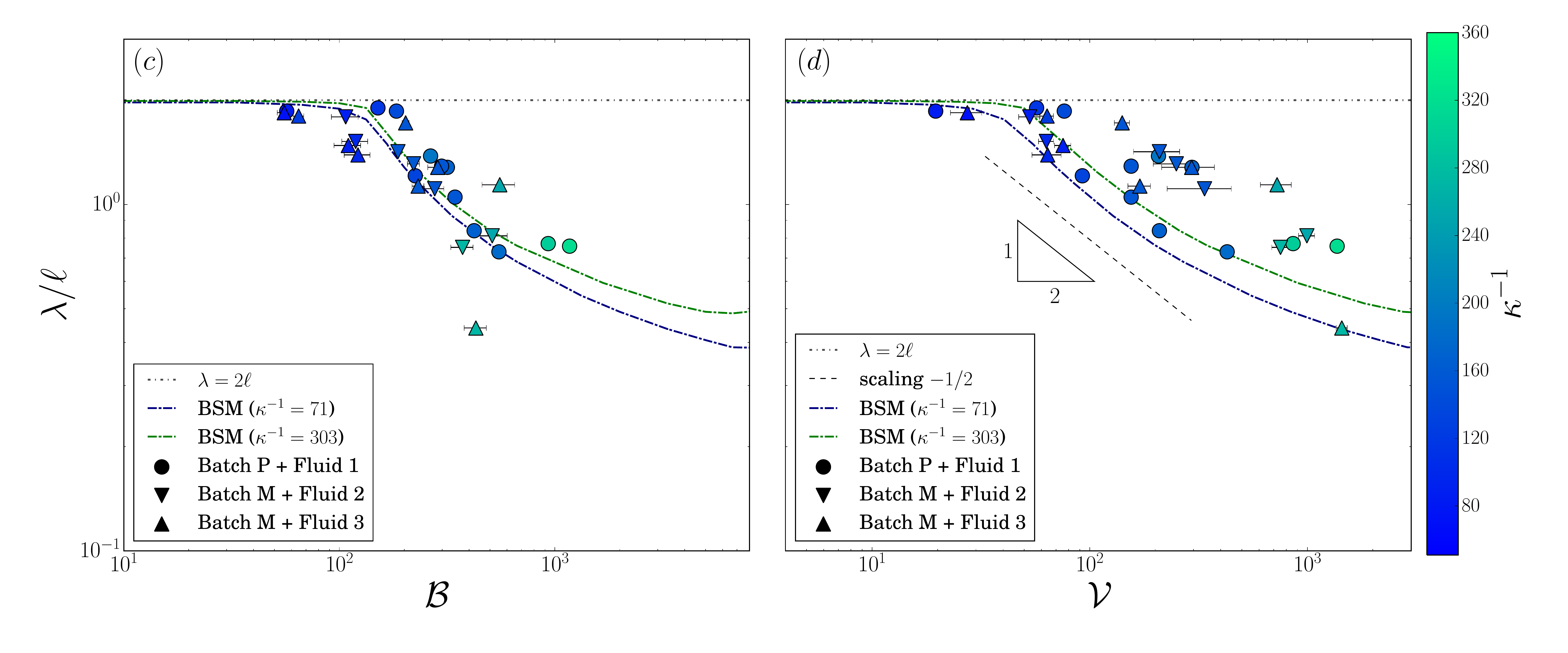}
\caption{Scaled maximum amplitude, $\delta/\ell$, (top) and end-to-end distance, $\lambda/\ell$, (bottom) as functions of $\mathcal{B}$ (left), graphs (a) and (c), and $\mathcal{V}$ (right), graphs (b) and (d), for the experimental, analytical (slender-body model), and numerical (bead-spring model noted BSM) results.}
\label{fig:deltalambda}
\end{figure}

Comparison of the experimental, analytical (slender-body model), and numerical (bead-spring model) results for the scaled maximum amplitude, $\delta/\ell$, and the end-to-end distance, $\lambda/\ell$, are plotted as functions of $\mathcal{B}$ and $\mathcal{V}$ in Fig.\,\ref{fig:deltalambda}. There is a reasonable agreement between the experimental data and numerical simulations. The experimental data present some scatters which reflect the experimental difficulties mentioned in Sec.\,\ref{sec:exp}. It is interesting to note that the experiments using the less viscous fluid 3, i.e. for which the Reynolds number is larger, do not present a notable different trend. This finding may be expected as the finite Reynolds corrections to the slender body theory only affect the order $1/(\ln \kappa^{-1})^2$ term in the drag and not the leading term \cite{KhayatCox1989}. Therefore, for the long filaments considered here, these finite Reynolds number corrections may not produce a significant effect. At small $\mathcal{B}$, the maximum deflection $\delta$ increases linearly with $\mathcal{B}$, in agreement with the small deformation limit given by Eq.\,(\ref{deltacyl}) of the slender-body theory \cite{Xu:1994}, see Fig.\,\ref{fig:deltalambda}\,(a). The amplitude depends on the aspect ratio, which varies between 80 and 300 as indicated by the color bar; larger aspect ratios exhibit smaller deflections, in agreement with Eq.\,(\ref{deltacyl}). For $\mathcal{B}\gtrsim 400$, the amplitude tends to saturate at $\delta \simeq 0.85 \ell$, independently of aspect ratio. It continues to increase with increasing $\mathcal{B}$ but at a very slow rate. This slow saturation of the amplitude cannot be captured by the slender body model but is predicted by the bead spring model in good agreement with the experimental observations.  The end-to-end distance $\lambda$ decreases with increasing $\mathcal{B}$ as the filament adopts a ``U'' shape, as shown in Fig.\,\ref{fig:deltalambda}\,(c). We observe the signature of the two limiting regimes; at low $\mathcal{B}$, the deformation is weak and $\lambda\simeq 2\ell$ while, at large $\mathcal{B}$, the deformation saturates and $\lambda$ decreases slowly towards $\lambda\simeq 0.3-0.4\ell$ with a small dependence on $\kappa^{-1}$. In addition, we identify here a third regime where $\lambda$ continuously evolves as the shape of the filament is modified. This reconfiguration regime is observed for $200<\mathcal{B}<500$ and corresponds to the intermediate profiles shown in Fig. \ref{fig:shapes}. The three regimes (weak deformation, reconfiguration, and saturation) are more conspicuous when plotting the data against the elasto-viscous number $\mathcal{V}$ in Fig.\,\ref{fig:deltalambda}\,(b),(d). The increase in data scatter is due to the uncertainty on the velocity used to estimate $\mathcal{V}$. For $\mathcal{V} \lesssim 30$, $\delta/\ell \propto \mathcal{V}$ and $\lambda/\ell\simeq 2$ (weak deformation). For $30 \lesssim \mathcal{V} \lesssim 400$, $\delta/\ell$ increases while $\lambda/\ell$ decreases; these evolutions are consistent with the scaling $\delta/\ell \propto \mathcal{V}^{1/2}$ and $\lambda/\ell \propto \mathcal{V}^{-1/2}$ derived for the reconfiguration regime in Sec.\,\ref{sec:Dimensional}. At larger $\mathcal{V}$, the deformation saturates toward a constant shape with $\delta/\ell\simeq 0.85$ and $\lambda$ decreases towards $\lambda/\ell \simeq 0.3-0.4$.

\subsection{\label{sec:velocity} Velocity}

\begin{figure}
\centering
\includegraphics[width=1\linewidth]{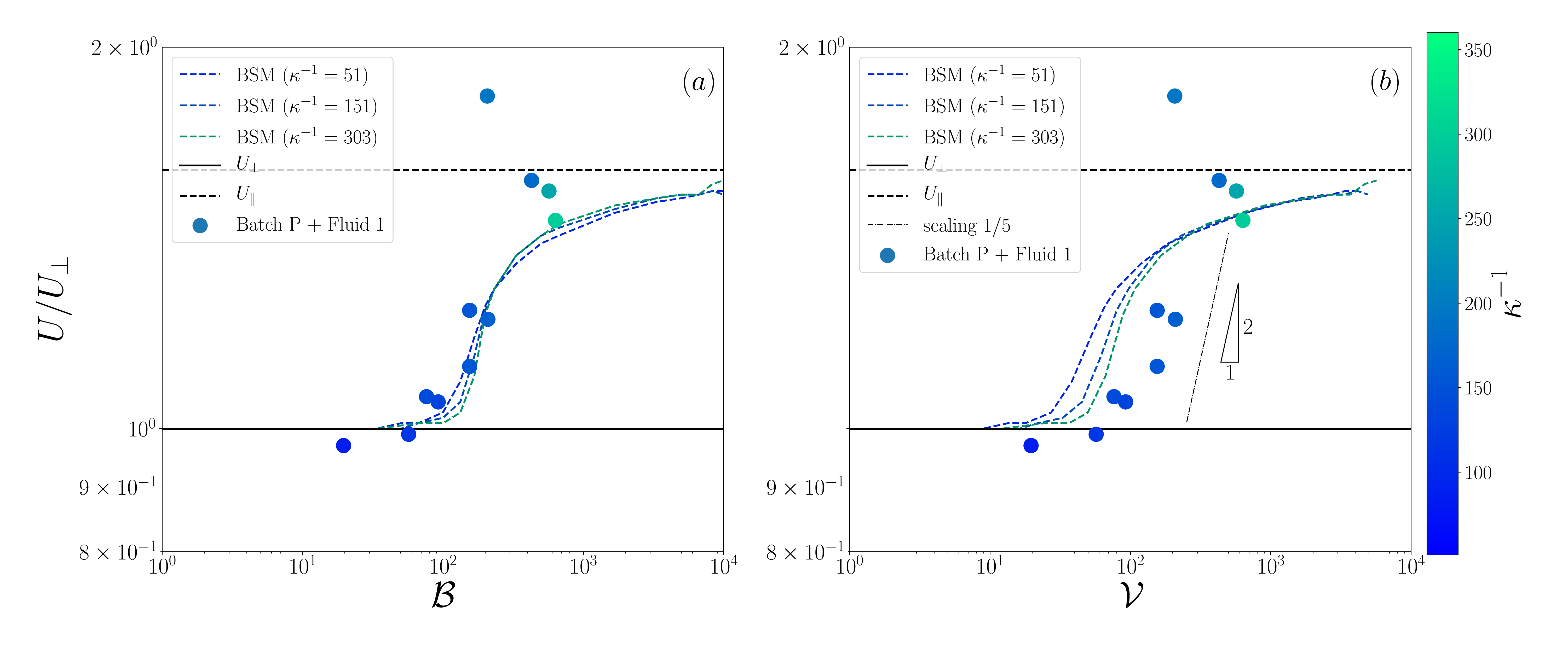}
\caption{Scaled velocity, $U/U_\perp$, versus $\mathcal{B}$ (left), graph (a), and $\mathcal{V}$ (right), graph (b), for the experimental and numerical (bead-spring model noted BSM) data.}
\label{fig:velocity}
\end{figure}

We measure the settling velocity of the filaments that we compare to the settling velocity of a rigid filament of same properties and dimensions, $U_{\perp}$ given by Eq.\,(\ref{eq:uperp}). These scaled velocities, $U/U_\perp$, are plotted as functions of $\mathcal{B}$ and $\mathcal{V}$ together with the numerical results coming from the bead-spring model in Fig.\,\ref{fig:velocity}. As discussed previously, the experimental measurements are affected by convection within the tank and by interactions with the  walls, resulting in a large scatter in the data; we thus only report data from batch P, in silicon-based fluid and a large tank, where these effects are lower. The complete sets of data are however given in the supplementary materials. We identify clearly the signature of the three regimes:  (i) for $\mathcal{B}<100$, $U\simeq U_{\perp}$, (ii) for $100<\mathcal{B}<1000$, the velocity then increases, and (iii) finally tends to saturate at a value close to the value $U=1.6 U_{\perp}$ corresponding to the settling velocity of a rigid vertical filament. Note that this value of $1.6 U_{\perp}$ overestimates the filament speed as the central part of the fiber remain horizontal and contributes to added drag on the filament. This marked evolution evidences the reconfiguration regime, where the shape of the filament evolves to reduce the drag. The drag force is no longer proportional to the velocity (which will give a constant velocity independent of $\mathcal{B}$), but rather can be expressed as $\propto U^{\alpha}$ with an exponent $\alpha<1$. The simple dimensional analysis proposed in Sec.\,\ref{sec:Dimensional} provides the scaling $U/U_{\perp} \sim \mathcal{V}^{1/2}$, in fair agreement with the data shown in Fig.\,\ref{fig:velocity}\,(b). 

\begin{figure}
\includegraphics[width=\linewidth]{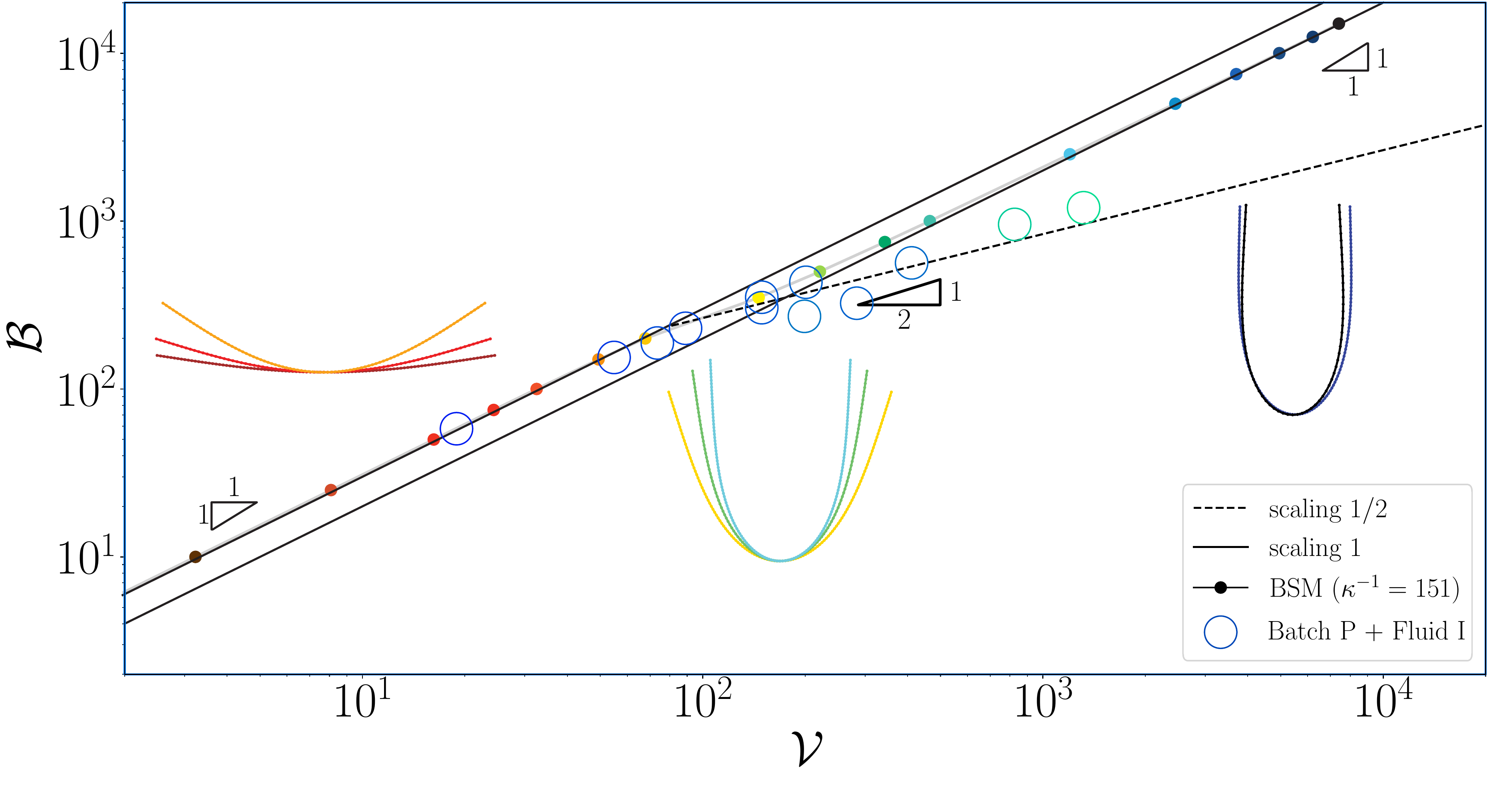}
\caption{Dimensionless drag, $\mathcal{B}$, versus dimensionless velocity, $\mathcal{V}$, for (a) the numerical (bead-spring model noted BSM) and experimental data of various aspect ratios and (b) for $\kappa^{-1}=151$ with the corresponding profiles.}
\label{fig:drag}
\end{figure}

Finally, the above results regarding the velocity of the filament can be summarized by plotting $\mathcal{B}$ against $\mathcal{V}$, i.e. the dimensionless drag as a function of a dimensionless velocity in Fig.\,\ref{fig:drag}. This graph evidences again the three regimes: (i) the weak (linear) deformation regime for which $\mathcal{B} \approx C_{\perp}\mathcal{V}$, (ii) the reconfiguration regime with $\mathcal{B} \sim \mathcal{V}^{\alpha}$ with an exponent $\alpha<1$, and (iii) the saturation for which
$\mathcal{B} \approx C_{\parallel}\mathcal{V}$. In contrast with high Reynolds number reconfiguration, in this low Reynolds number regime the lowest drag (achieved when the filament is perfectly aligned with the flow) has a finite value and differs only by a factor 1.6 (for the aspect ratios tested here) from the maximum drag experienced by a filament perpendicular to the flow. The reconfiguration regime is characterized by $\mathcal{B} \sim \mathcal{V}^{\alpha}$, and simple dimensional arguments give $\alpha= 1/2$; however, this scaling is left as soon as the filament shape approaches that of two vertical fibers, i.e. the reconfiguration regime is only a transient here, and only occurs in a limited range ($100\leq\mathcal{B}\leq1000$). We do not expect to recover perfectly the simple proposed scalings; however, they provide a qualitative insight in the mechanisms at play.

\section{\label{sec:conclusion} Conclusion}

In this work, we have presented a joint experimental, analytical, and numerical investigation of the equilibrium deformation of a flexible fiber settling in a quiescent viscous fluid. The major output of this study is the identification of three regimes having different signatures on the equilibrium configuration of the elastic filament. 

In the weak deformation regime, i.e. for small elasto-gravitational number, $\mathcal{B}$, or weak elasto-viscous number, $\mathcal{V}$, the filament adopts a ``V'' shape and its maximum deflection is linear in $\mathcal{B}$ as well as in $\mathcal{V}$ with a linear dependence on the inverse of the logarithm of the aspect ratio. In the large deformation (or saturation) regime, the filament takes a ``U'' shape (and can even reach a ``horseshoe'' shape) and both its maximum deflection and end-to-end-distance tend to saturate. These two regimes have been described in previous numerical work \cite{CosentinoLagomarsino:2005,Li:2013,Delmotte:2015} and are now further confirmed by the present observations. 

The important new finding of the present study is the existence of an intermediate regime of elastic reconfiguration. In the weak deformation regime, the drag of the filament becomes close to that of a rigid fiber settling perpendicular to the direction of gravity. In the large deformation regime, the drag is close to that of a rigid fiber settling in the parallel direction to gravity. In both cases, the drag is proportional to the velocity as is expected in Stokes flows. Conversely, in the intermediate reconfiguration regime, the filament deforms to adopt a shape with a smaller drag which is no longer proportional to the velocity but rather to the square root of the velocity, i.e. $\mathcal{B} \sim \mathcal{V}^{1/2}$. This crossover regime between the linear and saturation regimes, while anticipated in \cite{CosentinoLagomarsino:2005}, has been clearly identified through its different scaling behavior in the present work combining experiments and a simple bead-spring modeling of the filament.

\begin{acknowledgements}
This work was undertaken under the auspices of the ANR project `Collective Dynamics of Settling Particles In Turbulence' (ANR-12-BS09-0017-01), the `Laboratoire d'Excellence M\'ecanique et Complexit\'e' (ANR-11-LABX-0092), the `Initiative d'Excellence' A$^{\ast}$MIDEX (ANR-11-IDEX-0001-02), funded by the French Government `Investissements d'Avenir programme', the COST Action MP1305 `Flowing Matter', the ANR project `DeFHy' and the ERC Consolidator Grant `PaDyFlow' in the framework of the Horizon 2020 program under grant agreement no 682367. We thank E. de Langre for useful discussions.
\end{acknowledgements}

\end{document}